# Fe-As bond fluctuations in a double-well potential in LaFeAsO

Valentin G. Ivanov[1], Andrey A. Ivanov[1], Alexey P. Menushenkov[1], Boby Joseph[2,3] and Antonio Bianconi[1,3*]

**Abstract**  While for a long time iron based superconductors have been investigated looking at the average crystalline structure there is now growing interest on the divergence of local from the average structure. Here we use advanced data analysis of EXAFS, a fast and local probe, which has key advantages compared to crystallographic measurements, in unveiling the dynamics of the local structure. The results show that the Fe-As pair oscillates in the double-well potential both in undoped and in cobalt doped LaFeAsO crystals. The parameters and the characteristics of the double-well potential are determined by curve fits of the polarized EXAFS data. Temperature dependent Fe-As pair distribution and tunneling frequency are derived from selected model of two potential wells. We observe the softening of the Fe-As oscillations at the tetragonal to orthorhombic structural phase transition at 150 K in LaFeAsO. A weak broad softening is observed while decreasing the temperature from 70 to 40 K in superconducting $LaFe_{0.89}Co_{0.11}AsO$ which could be correlated with the onset of short range charge density wave order.

---

*corresponding author: Antonio Bianconi
E-mail: antonio.bianconi@ricmass.eu

[1]*National Research Nuclear University MEPhI (Moscow Engineering Physics Institute), Kashirskoe sh. 31, 115409 Moscow, Russia*

[2]*Elettra, Sincrotrone Trieste, Strada Statale 14, Km 163.5, Basovizza, 34149 Trieste, Italy*

[3]*Rome International Centre for Material Science Superstripes, RICMASS, via dei Sabelli 119A, 00185 Rome, Italy*

# 1 Introduction

There is growing evidence that the divergence of local structure from average structure is a generic feature of high temperature superconductors due to inhomogeneity of charge, lattice, spin and superconductivity short range order [1-7]. This complex scenario was proposed to be characteristic of a multiband correlated electronic system where one Fermi surface is in the proximity of Lifshitz transition [7-9]. This scenario is characterized by multiscale inhomogeneity going from micron scale to atomic scale. [1-6]. While this inhomogeneity was identified in cuprates [1-4] and chalcogenides [5, 6] there is lack of information on local fluctuations in RFeAsO (R = La, Nd, Sm), called 1111 systems [10-12]. Lattice and charge inhomogeneous fluctuations have been predicted theoretically [13]. However, experimental investigation of intrinsic local lattice fluctuations needs single crystals of high quality which were not available so far. Recently 1111 single crystals suitable for experimental investigations have been grown [14,15] and diffraction [16] as well as polarized extended x-ray-absorption fine structure (EXAFS) experiments [17] have been reported.

The EXAFS structure in X-ray absorption spectra provides information on the local structure around a selected atomic species with a femtoseconds measuring time scale. Therefore it provides the atomic pair radial distribution function between selected atomic species probing instantaneous bond length distribution without time averaging [18-23]. Using this technique polarons [19], the striped short range charge density wave accompanied by periodic lattice distortions [20, 21], the isotope effect on striped structure [22], different from isotope effect on Tc [23], have\ been determined in cuprates.

Here we report the analysis of polarized As K-edge extended x-ray-absorption fine-structure (EXAFS) measurements of undoped and cobalt doped LaFeAsO performed to unveil the local fast Fe-As zero point fluctuations in a double-well potential. This approach was successfully applied to hole doped cuprates [24-27], $Ba_{1-x}K_xBiO_3$ and electron doped cuprates [28-30]. Polarized As K-edge EXAFS in 1111 systems gives instantaneous pair distribution function between As and neighboring Fe atoms. In fact EXAFS is a probe of interference pattern due to the scattering of As emitted photoelectron by neighboring Fe atoms at time scale of $10^{-15}$ s and spatial scale of 0.5 nm. The parameters of double-well potential are determined by optimization of fits of the experimental EXAFS data. This approach ideally suits for anharmonic systems because the suggested potential can be a very close approximation to the real one and it also offers the advantage of providing dynamical information. The structural phase transition at 150 K in the undoped system appears to involve the softening of double-well potential in which the distance between two wells and the barrier height have both decreased. This softer structure is present at fluctuation region below phase transition temperature. We discuss how the double-well is affected by the substitution of cobalt for iron, providing further evidence for sensitivity of factors that influence superconductivity to atoms displacement.

# 2 Results and Discussion

The EXAFS experimental data and their Fourier transform (FT) extracted from As K-edge EXAFS in the whole temperature range are presented in Fig. 1. Fourier transform magnitudes of As $K$-edge (k range 3.5-17 Å$^{-1}$) EXAFS oscillations, measured on LaFe$_{1-x}$Co$_x$AsO (x=0 and 0.11) single crystals at representative temperatures, providing partial atomic distribution around As atom are shown in the upper and lower left panels of Fig. 1.

The photoelectron excited from As atom at the absorption As $K$-edge is scattered by Fe near neighbors at distance of about 2.4 Å as indicated by Fourier transform peak in Fig. 1. The contributions of La and O atoms neighbors of As are mixed in the Fourier tranform giving peaks between 2.6 and 4.0 Å. Therefore the contributions of As-Fe bonds are well separated from other atomic contributions and the quantitative information on the distribution of instantaneous Fe-As bonds can be easily extracted as shown in the inset of Fig. 1. The filtered EXAFS oscillations corresponding to Fe-As contributions (shown in insets of Fig. 1 for representative temperatures) exhibit a smooth damping with increasing wavevector. Usually *K*-edge EXAFS of powder samples have been analyzed using a single shell fit to extract information on the Fe-As bond correlations. For this case Fe-As distance and related mean square relative displacement (MSRD), describing correlated Debye-Waller factor ($\sigma^2$), were determined in conventional least squares modelling, using phase and amplitude factors [17].

The EXAFS spectra have been collected between 12 and 300 K. A structural transition from tetragonal to orthorhombic phase in LaFeAsO is known to occur at around 150K [11,12]. This transition for our sample is characterized by the orthorhombic distortion occuring below 154.5K and the onset of magnetic order around 140 K. Both these transitions are absent in Co doped sample, which has superconducting transition around 10 K. For the parent compound Fe-As MSRD shows abrupt change around 150 K [17]. No such change is seen for Co doped sample. Clearly, the anomalies seen in the Fe-As MSRD are related to the phase transitions in LaFeAsO. Such anomalies were not reported in the earlier EXAFS studies of polycrystalline powder samples.

Here the investigation of the low temperature anomaly in local lattice fluctuations was performed by EXAFS data analysis in the framework of double-well potential approach following the methodology developed elsewhere [28]. It was assumed that in iron-based superconductors Fe atoms could also dynamically oscillate near As atoms in double-well potential that is possible to detect by EXAFS. Polarized EXAFS enables to probe mainly one type of Fe-As bonds oriented along the electric field direction of the incident photon beam.

The double-well potential was constructed of two parabolic functions connected continuously:

$$U(r) = \frac{a}{2}(r-x)^2 \, lstep\left(r, \frac{z + x\sqrt{\frac{a}{b}}}{1 + \sqrt{\frac{a}{b}}}\right) + \frac{b}{2}(r-z)^2 \, rstep\left(r, \frac{z + x\sqrt{\frac{a}{b}}}{1 + \sqrt{\frac{a}{b}}}\right), \qquad (1)$$

where $lstep(r, A) = \begin{cases} 1, r < A \\ 0, r > A \end{cases}$ and $rstep(r, B) = \begin{cases} 0, r < B \\ 1, r > B \end{cases}$ — step-functions;

*a, b* are the force constants of the first and the second potential wells, which have the minima at *x* and *z*, respectively. Therefore, the splitting of the Fe-As bond distance in the double-well is given by $\Delta = z \text{-} x$.

The fit of the experimental data was performed by using the VIPER program [31] in 2 steps. Initially, all 4 parameters of the potential (*a, b, x, z*) were varied, starting from $a = b = 1.0 \times 10^6$ K/Å$^2$, $x = 2.35$ Å, $z = 2.45$ Å, i.e., $\Delta = 0.1$ Å. The final refinement has been done by varying only wells' positions *x, z*. Thus, force constants *a, b* of the double-well potential were fixed at their average values found during the initial fit (insulator: $a = b = 1.63 \times 10^6$ K/Å$^2$; superconductor: $a = b = 1.14 \times 10^6$ K/Å$^2$). The best fit is obtained with a double-well potentials with two nearly equally populated wells of Fe-As split by 0.1 Å with a tunneling frequency in the range between 3 and 7 THz.

Double-well potentials with $E_0$, $E_1$ energy levels and Pair Distribution Functions (PDF) are shown on Fig. 2. Panel (a) and panel (b) of Fig.2 show potential double-well and the bimodal Fe-As pair distribution function at 100 K and 180K for undoped material above and below the structural phase transition. Panel (c) and panel (d) of Fig.2 show the potential well and the PDF of the superconducting material at 25K and 70K. The important changes of the double-well potential take place while going through the temperature variation.

The barrier form modification leads to the change in the Fe atom frequency of tunneling ($\omega = E_1 \text{-} E_0$) between wells, presented on Fig. 3. Panels (a) and (b) of Fig.3 show the tunneling frequency and the distance between wells in the insulating material. We see that the distance between wells decreases from 0.105 Å at high temperatures to 0.096 Å at 150K with sharp minimum of 0.089 Å at the structural transition. The tunneling frequency increases from 3.5 THz up to 4.5 THz below the structural transition with the sharp maximum at the transition reaching 6.5THz. This plot shows that the increase of tunneling frequency occurs at 150 K for insulator LaFeAsO which is due to the decrease of the distance $\Delta = z - x$ between wells. The corresponding increase of the probability to find Fe atom in the barrier region is also well seen on the representative plots of PDF (Fig. 2).

Panel (d) of Fig. 2 shows for superconductor LaFe$_{1-x}$Co$_x$AsO (x=0.11) a similar merging of the double-well going from 70 K to 25 K with the decrease of the potential barrier shown in panel (c). In fact the distance between the minima of the double-well decreases from 0.103 Å at 70 K to 0.094 Å at 25 K with smooth behavior (see Fig. 3(d)). Similarly, as shown in Fig. 3(c), the tunneling frequency increases with a smooth behavior from 4.5 THz at 70 K to 6 THz at 25 K.

**3 Conclusions**

In conclusion, we present the results of temperature dependent EXAFS study on LaFe$_{1-x}$Co$_x$AsO (x=0 and 0.11) single crystals at As $K$-edge. Fe-As bond was analyzed using double-well potential model. The results show that this model gives more precise description than single well one. It was demonstrated that Fe-As pair oscillates in double-well potential both in undoped and in cobalt doped LaFeAsO crystals. The best fit is obtained with a double-well potentials with two nearly equally populated wells of Fe-As split by 0.1 Å with a tunneling frequency in the range between 3 and 7 THz. We observe the softening of the Fe-As oscillations at the tetragonal to orthorhombic structural phase transition at 150 K in LaFeAsO. A weak broad softening is observed while decreasing the temperature from 70 to 40 K in superconducting LaFe$_{0.89}$Co$_{0.11}$AsO which could be correlated with the onset of short range charge density wave order. Similar effects involving copper - oxygen atoms have been observed in hole and electron doped cuprates. These results have been interpreted in terms of polarons, short range charge density waves and anharmonic lattice dynamics which are difficult to be unveiled by methods averaging the response in time and space.

**Acknowledgements** All authors thank ESRF for beam time allocation, the staff of BM-23 beam line, Nicola Poccia and Alessandro Ricci for running the experiment; V.G.I., A.A.I, and A.P.M. thank the Russian Science Foundation (project 14-22-00098) for financial support; A.B. acknowledge financial support of superstripes-onlus.

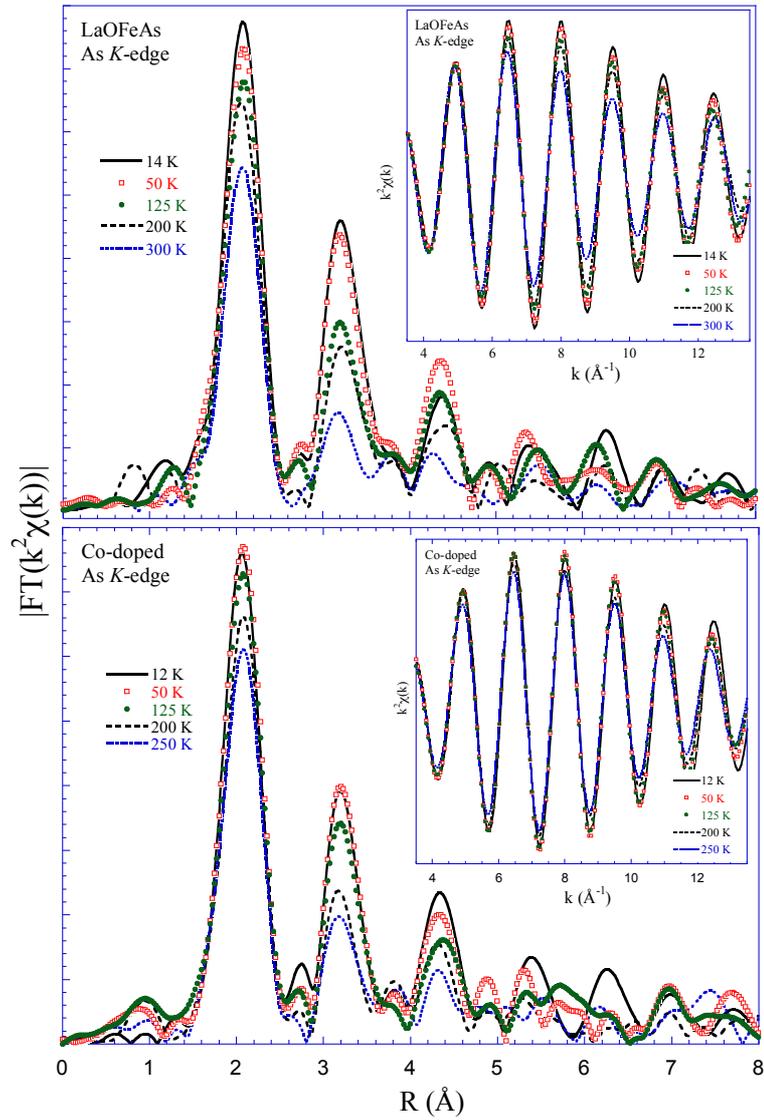

**Figure 1**: Fourier transform (FT) magnitudes of the As *K*-edge EXAFS oscillations at different representative temperatures for the LaFe$_{1-x}$Co$_x$AsO (x=0 and 0.11) single crystals. FTs are not corrected for the phase shifts, and represent raw experimental data. Inset shows the filtered EXAFS corresponding to the first shell.

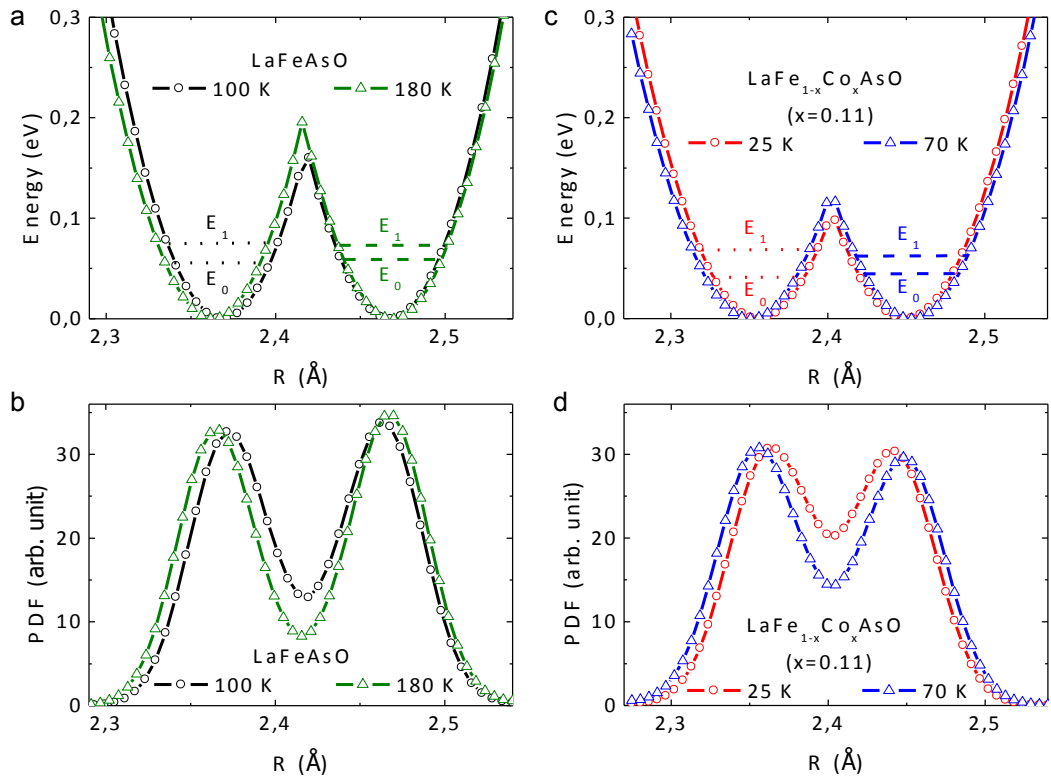

**Figure 2**: Double-well potentials with $E_0$, $E_1$ energy levels and PDF for insulator LaFeAsO (left panel) and superconductor LaFe$_{1-x}$Co$_x$AsO (x=0.11) (right panel).

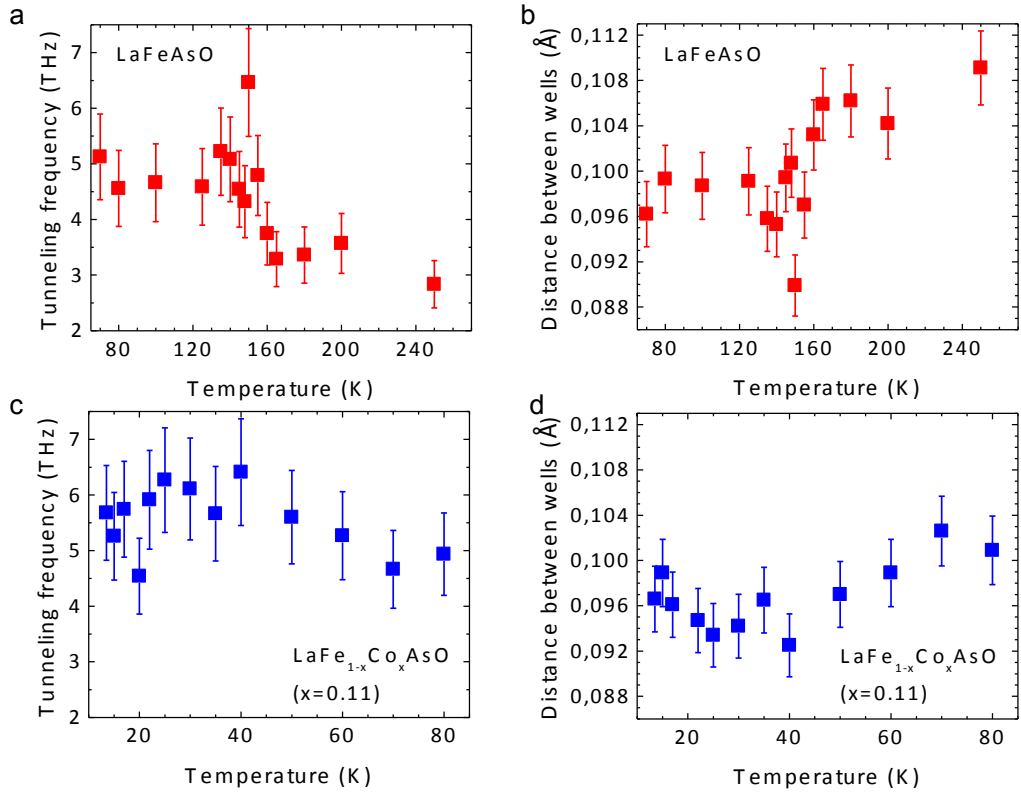

**Figure 3**: Temperature dependent frequency of Fe atom tunneling through the double-well potential barrier (left panel) and the corresponding distance between wells (right panel) for insulator LaFeAsO and superconductor LaFe$_{1-x}$Co$_x$AsO (x=0.11).